\documentclass[apjl]{emulateapj}
\usepackage{apjfonts}

\newcommand{\magsec}{mag arcsec$^{-2}$}

\newcommand{\eg}{e.g.,\ }
\newcommand{\ie}{i.e.,\ }
\newcommand{\etal}{et~al.\ }

\newcommand{\muv}{$\mu_{V}$}
\newcommand{\mub}{$\mu_{B}$}
\newcommand{\Bp}{$B'$}
\newcommand{\B}{$B$}
\newcommand{\M}{$M$}
\newcommand{\V}{$V$}
\newcommand{\BV}{$B-V$}
\newcommand{\ltsima}{$\; \buildrel < \over \sim \;$}
\newcommand{\simlt}{\lower.5ex\hbox{\ltsima}}
\newcommand{\gtsima}{$\; \buildrel > \over \sim \;$}
\newcommand{\simgt}{\lower.5ex\hbox{\gtsima}}
\newcommand{\ppix}{pix$^{-1}$}
\newcommand{\by}{$\times$}

\begin{document}

\title{Optical Colors of Intracluster Light in the Virgo Cluster Core}
 
\author{Craig S. Rudick\altaffilmark{1},
  J. Christopher Mihos\altaffilmark{1},
  Paul Harding\altaffilmark{1},
  John J. Feldmeier\altaffilmark{2},
  Steven Janowiecki\altaffilmark{3},
  and Heather L. Morrison\altaffilmark{1}}

\email{csr10@\-case.\-edu}

\altaffiltext{1}{Department of Astronomy, Case Western Reserve
  University, 10900 Euclid Ave, Cleveland, OH 44106, USA}
\altaffiltext{2}{Department of Physics and Astronomy, Youngstown State
  University, Youngstown, OH 44555, USA}
\altaffiltext{3}{Department of Astronomy, Indiana University, 727 East 3rd Street, Bloomington, IN 47405, USA}

\begin{abstract}
We continue our deep optical imaging survey of the Virgo cluster using
the CWRU Burrell Schmidt telescope by presenting \B-band surface
photometry of the core of the Virgo cluster in order to study the
cluster's intracluster light (ICL).  We find ICL features down to
\mub$\approx29$ \magsec, confirming the results of Mihos \etal
(2005), who saw a vast web of low-surface brightness streams, arcs,
plumes, and diffuse light in the Virgo cluster core using \V-band
imaging.  By combining these two data sets, we are able to measure the
optical colors of many of the cluster's low-surface brightness
features.  While much of our imaging area is contaminated by galactic
cirrus, the cluster core near the cD galaxy, M87, is unobscured.  We
trace the color profile of M87 out to over 2000\arcsec, and find a
blueing trend with radius, continuing out to the largest radii.
Moreover, we have measured the colors of several ICL features which
extend beyond M87's outermost reaches and find that they have similar
colors to the M87's halo itself, \BV$\approx0.8$.  The common colors
of these features suggest that the extended outer envelopes of cD
galaxies, such as M87, may be formed from similar streams, created by
tidal interactions within the cluster, that have since dissolved into
a smooth background in the cluster potential.
\end{abstract}

\keywords{galaxies: clusters: individual (Virgo) --- galaxies:
  individual (M87) --- galaxies: interactions --- galaxies:
  photometry}

\section{Introduction}
Massive galaxy clusters are known to contain a population of stars
which reside outside of any of the cluster's galaxies, often referred
to as intracluster light or ICL.  ICL features typically have
extremely faint surface brightnesses of $<1\%$ of the brightness of
the night sky, making their study extremely difficult.  While the
first indications of the existence of ICL came from observations by
Zwicky (1951), only with the advent of modern CCD technologies have
detailed studies of these stars been made possible (\eg Uson \etal
1991; V{\'i}lchez-G{\'o}mez et al. 1994; Bernstein \etal 1995; Gregg
\& West 1998; Trentham \& Mobasher 1998).

The most straightforward method of detecting the ICL is through deep
broadband imaging at optical wavelengths (\eg Uson \etal 1991;
V{\'i}lchez-G{\'o}mez et al. 1994; Trentham \& Mobasher 1998;
Feldmeier \etal 2002, 2004a; Mihos \etal 2005, hereafter referred to
as M05; Gonzalez \etal 2005; Krick \& Bernstein 2007).  With such
imaging, we can not only measure the luminosity of ICL the component,
but we can study its spatial distribution and detect individual ICL
features, such as streams, arcs, and plumes (\eg Gregg \& West 1998;
Trentham \& Mobasher 1998; Calc{\'a}neo-Rold{\'a}n \etal 2000; White
\etal 2003; M05; Krick \etal 2006; Yagi \etal 2007) as well as any
large-scale diffuse components (\eg Gonzalez \etal 2000; Feldmeier
\etal 2002, 2004a; Adami \etal 2005; Zibetti \etal 2005; Patel \etal
2006; Krick \& Bernstein 2007; Pierini \etal 2008; Da Rocha \etal
2008).  ICL has also been detected using discrete stellar tracers (\eg
Ferguson \etal 1998; Feldmeier \etal 1998; Durrell \etal 2002; Gal-Yam
\etal 2003; Arnaboldi \etal 2004; Feldmeier \etal 2004b; Gerhard \etal
2005; Aguerri \etal 2005; Neill \etal 2005; Maoz \etal 2005; Williams
\etal 2007; Castro-Rodrigu{\'e}z \etal 2009; McGee \& Balogh 2010),
although these methods often lack the ability to detect individual ICL
features.

Intracluster light is thought to form primarily by the tidal stripping
of stars as galaxies interact and merge during the hierarchical
accretion history of the cluster, causing the fraction of the
cluster's luminosity found in the ICL to increase as it evolves (\eg
Napolitano \etal 2003; Murante \etal 2004; Willman \etal 2004; Rudick
\etal 2006; Monaco \etal 2006; Conroy \etal 2007; Murante \etal 2007;
Purcell \etal 2007; Yan \etal 2009; Baria \etal 2009).  Numerous
mechanisms for generating the ICL have been proposed, including the
infall of groups into the cluster potential (Willman \etal 2004;
Rudick \etal 2006), high speed encounters within the cluster (Moore
\etal 1996; Gnedin 2003), and galactic mergers during the buildup of
the massive central galaxy (Murante \etal 2007; Conroy \etal 2007).
All of these processes are likely to be occurring simultaneously and
each will create distinct observable signatures (Rudick \etal 2009).
Thus, the formation of ICL is intimately linked to the dynamical
history of the cluster, and the observable features of the ICL should
contain a great deal of information about the evolutionary processes
which have shaped both the cluster and its constituent galaxies.

While ICL is generally thought of as stellar material found
\emph{outside} of any individual galaxy, in practice galaxies have no
well-defined edge (Abadi \etal 2006).  Thus, the distinction between
the ICL and the outer luminosity profiles of cluster galaxies, which
display similar surface brightnesses in broadband imaging, is
difficult and somewhat arbitrary (M05).  In fact, simulations of the
formation of clusters' most massive elliptical galaxies have shown
that their extended stellar profiles form through similar merger and
tidal stripping mechanisms as the more diffuse ICL (Dubinski \etal
1998; Monaco \etal 2006; Conroy \etal 2007; Murante \etal 2007;
Ruszkowski \& Springel 2009).  We therefore prefer to use
\emph{intracluster light} as a qualitative description of cluster
luminosity at low surface brightness, which may refer to individual
tidal streams, extreme galactic outskirts, or any large-scale diffuse
luminosity component, and which are all likely products of tidal
stripping and disruption of galaxies during the dynamical evolution of
the cluster.

In addition to the quantity and morphology of the ICL, knowledge of
the underlying stellar populations also provides a vital tool for
understanding its formation.  Giant elliptical galaxies are known to
display radial color gradients, whereby the color index decreases, or
becomes bluer, with increasing radius (\eg Pettit 1954; de Vaucouleurs
1961; Carter \& Dixon 1978; Strom \& Strom 1978; Davis \etal 1985;
Vader \etal 1988; Goudfrooij \etal 1994; Bernardi \etal 2003;
Cantiello \etal 2005; Liu \etal 2005), primarily due to gradients in
the stellar metallicities (\eg Spinrad \etal 1972; Strom \etal 1976;
Baum \etal 1986; Carollo \etal 1993; Tamura \etal 2000;
S{\'a}nchez-Bl{\'a}zquez \etal 2007; Rawle \etal 2010).  Just as the
strength of these gradients is an important clue in reconstructing the
formation history of elliptical galaxies, the ages and metallicities
of the intracluster stars will be highly dependent on the progenitor
galaxies from which they were stripped.  The results of Sommer-Larsen
\etal (2005) show that the metallicity of the ICL is expected to be on
average similar to that of the outer envelope of the cD galaxy, while
Purcell \etal (2008) show that individual ICL streams formed by recent
interactions should be more metal rich than the surrounding diffuse
component.  Murante \etal (2004) have shown that the diffuse ICL
stellar population is expected to be older, on average, than the
galactic stars.

Observationally, it is very difficult to break the age-metallicity
degeneracy and make precise determinations of either quantity.
Although Williams \etal (2007) used HST ACS imaging of red giant
branch stars in the Virgo cluster to do so, such observations are only
possible in the most nearby systems and span very small fields of
view.  Their work found that the ICL is composed of stars with a wide
variety of ages and metallicities, but that the dominant component is
old ($\simgt 10$ Gyr) and moderately metal poor ([M/H]$\simlt -1.0$).
A more common technique is to instead use broadband optical colors in
order to compare the ICL stellar populations to those of galactic
stars.  Most studies have found results consistent with Williams \etal
(2007), where the average ICL color is similar to that of the outer
halo of the cluster's brightest galaxy (\eg Zibetti \etal 2005;
Pierini \etal 2008; da Rocha \etal 2008), which due to radial color
gradients is bluer than the galaxy's interior regions.  However, a
number of clusters have been observed to have an ICL component with
significantly redder colors, more similar to the brightest galaxy's
interior (Krick \& Bernstein 2007; da Rocha \& Mendes de Oliveira
2005; Gonzalez \etal 2000).

As part of our ongoing deep imaging survey of the Virgo cluster using
Case Western Reserve University's Burrell Schmidt telescope, this
paper presents deep imaging of the Virgo cluster core in the \B-band.
By combining this data with the \V-band results previously published
in M05, we are able to measure the optical colors of Virgo's
intracluster light.  A detailed description of our data acquisition
and reduction techniques is given in Section 2.  Our \B-band image is
presented in Section 3, while Section 4 combines the two images in
order to measure the colors of ICL features.  Section 5 includes a
summary of our results and a discussion of our interpretations.
Finally, detailed error models for our photometric measurements can be
found in the Appendix.

\section{Observations and Data Reduction}
\subsection{Observations}\label{sec:observations}
Our imaging data was collected on dark photometric nights on three
separate observing runs spanning January through March of 2009, using
the 0.6m CWRU Burrell Schmidt Telescope located at Kitt Peak National
Observatory.  The Burrell Schmidt's closed-tube optical design makes
it ideal for wide-field, deep optical imaging.  Using a
4096$\times$4096 pixel CCD read out of 4 amplifiers simultaneously,
the 1.45\arcsec\ pixel scale results in a field of view which exceeds
2.5 square degrees.  Further details of the CWRU Burrell Schmidt's
optical system can be found in Nassau (1945) and Slater \etal (2009).

The observations described in M05 were made using the Washington \M\
filter, converted to Johnson \V\ magnitudes.  To measure the colors of
the ICL features seen in those observations, we could have selected to
re-observe the area in either a bluer or redder filter.  However, at
redder wavelengths, emission lines in the atmosphere cause the night
sky to vary considerably in brightness on timescales of minutes
(Feldmeier \etal 2002 and references therein), which would severely
hamper our ability to accurately flat field and sky-subtract our data.
While both our targets and the sky are fainter in the blue, requiring
longer observing times, the stability of the night sky emission made
the selection of a blue filter the natural choice.

Because the Washington \M\ filter is somewhat shifted blue-ward of the
\V\ filter, and to ensure adequate wavelength separation between our
two filters, for these observations we used a custom-designed filter,
referred to as \Bp, with an effective wavelength $\approx200 \AA$
shorter than the standard Johnson \B\ filter.  All surface brightness
measurements, however, have been transformed to the Johnson \B\
system.

Our observational and data reduction techniques are based on those
described in Morrison \etal (1997), Feldmeier \etal (2002), M05, and
Slater \etal (2009).  Because of the diffuse, extremely faint nature
of the ICL features we are measuring ($0.1-1\%$\ of the brightness of
the night sky), generating a flat-field which exceeds this precision
over the entire field of view is of critical importance.  We therefore
constructed a night-sky flat from pointings of pre-selected sky
fields, chosen to be offset from our target field by $\approx0.25-1$\
hours in right ascension and less than $\approx4$\degr\ in
declination, while containing relatively low stellar density and no
particularly bright stars.  Our observing schedule was roughly evenly
divided between imaging our target and the offset sky fields.  A
typical observation cycle consisted of 2 sky images offset west, 4
images on the target field, and 2 more sky images offset east,
repeated throughout the night.  By minimizing the motion of the
telescope during the cycle, this pattern reduced any hysteresis
effects caused by flexure of the optical system and ensured that the
sky frames were taken as close in time and position as possible to the
target images, under similar atmospheric and telescope conditions.
When imaging the target field, individual pointings were intentionally
offset in a dither pattern from the field center by as much as
0.75\degr, or half the size of the field of view, in order to reduce
systematic effects which may result from consistently imaging objects
with the same area of the CCD.  All sky and target field images were
1200 second integrations, yielding a sky brightness of $\approx 750$
ADU.

\subsubsection{Sources of Diffuse Light}
Although the scientific aim of our survey is to image the intracluster
light in the Virgo galaxy cluster, we are, of course, sensitive to all
sources of low-surface brightness diffuse light.  In practice, we
detect two major sources of diffuse light: ICL in the Virgo cluster
and light scattered by galactic cirrus within our own Galaxy (\eg
Sandage 1978; Guhathakurta \& Tyson 1989; Witt \etal 2008).  For the
purposes of image processing and data reduction, all diffuse light
sources are equivalent and we make no distinction between luminosity
from galactic or extra-galactic sources.  For scientific analyses,
however, the source of the diffuse light can be critical to the
interpretation of our results, and the impact of galactic cirrus is
discussed in Section \ref{sec:cirrus}.

\subsection{Image Pre-Processing, Photometric Solutions, and Data Quality Cuts}
Our image processing procedure began by subtracting a nightly bias
frame, applying a CCD overscan region correction, and removing
amplifier crosstalk from all images, each in the usual manner
(hereafter, these images are referred to as pre-processed).  The
images were then flattened using a preliminary flat field generated by
a simple median combination of all of our sky frames scaled by their
mode, using the \emph{imcombine} task from the IRAF software package
\footnote{IRAF is distributed by the National Optical Astronomy
Observatory, which is operated by the Association of Universities for
Research in Astronomy (AURA) under cooperative agreement with the
National Science Foundation.}.  Although the input sky images were not
put through the rigorous screening and processing required to create a
final flat field (see Section \ref{sec:flatfield}), pixel values in this
preliminary flat field, created from 111 individual images, differed
from the final version by typically $<0.5\%$.  An astrometric solution
for each image was calculated using the Astrometry.net software
package (Lang \etal 2010), and an airmass correction was applied.

Photometric zeropoints for all images were then calculated by
comparing the stellar fluxes, as measured using IRAF's \emph{daophot}
routines, to those found in the Sloan Digital Sky Survey DR7 catalog
(Abazajian \etal 2009), transformed to Johnson \B\ magnitudes (Ivezic
\etal 2007).  From this analysis, we calculated a \BV\ color term of
0.05 mag to transform our data to the Johnson \B\ system.  For each
night of observations we calculated and subtracted a nightly
zeropoint, taken to be the mean of all zeropoints from that night.
After subtracting this nightly zeropoint, all images were set to a
common photometric zeropoint.  For a typical \BV\ color of 1.0, 1 ADU
corresponds to 29.6 \magsec.

A data quality cut was made by excluding images from nights found to
have highly variable photometric zeropoints, indicative of light cloud
cover or other non-photometric conditions.  We calculated the mean and
scatter in zeropoint for all images and excluded any data from nights
on which the photometric zeropoint varied by $>2\sigma$\ from the
mean, or approximately 0.02 mag.  This quantitative measure of
photometric atmospheric conditions agreed remarkably well with
observer notes from the nightly observation logs.

\subsection{Flat Field}\label{sec:flatfield}
Our flat field was constructed using the offset sky images described
in Section \ref{sec:observations}.  The primary motivation for using this
sky-flat method is that we expect the detector response to be
wavelength-dependent.  By using the night sky as our uniform
illumination source, we ensure that the flat field is measured at the
same effective wavelength as our target sources.  Because the diffuse
light sources we are measuring are $<1\%$ of the brightness of the
night sky, we are dominated by sky photons even in our target frames.
Moreover, we expect the ICL itself to be similar in color to the night
sky (Taylor \etal 2004; Sommer-Larsen \etal 2005), further motivating
the use of the night sky as our uniform illumination source.

The first step in creating the final flat field was to use IRAF's
\emph{objmask} task to find and mask all objects in each image,
including stars, galaxies, satellite trails, and any other bright
features.  Because our goal is to measure a uniform illumination
pattern as well as possible, we used an aggressive masking technique
whereby we ran \emph{objmask} twice, first on the original image and
again after masking out any objects found in the first run.  Once the
object masks were applied, we median-binned the images in
128$\times$128 pixel groups, and visually inspected each frame to
check the image quality and search for and mask any remaining diffuse,
non-sky features.  All images had some features which were masked by
hand, including the extended wings of bright stars, internal
reflections, and flares from stars near the edge of the field of view.
A number of images were discarded due to the presence of strong
galactic cirrus, a particularly bright star on the image, or other
features which created a large-scale illumination pattern.  After this
quality cut we were left with 84 night sky images with which to create
the final flat field.

Due to variations in the atmospheric conditions of each observation,
such as airglow, transparency, airmass, etc., the mean flux level of
the night sky varies from image to image, typically in the range of
650-850 ADU, or \mub=22.6-22.3 \magsec.  However these effects not
only cause the mean sky level to vary, they are also the cause of an
illumination gradient over the field of view of each image, typically
in the range of 5-10 ADU, or $\approx1\%$ of the sky flux across the
frame.  In order to account for these atmospheric effects and create a
flat field representing a uniform illumination pattern, we employed an
iterative process to combine the night sky images into our final flat
field, following Morrison \etal (1997).  We began by flattening the
pre-processed sky images using the preliminary flat field and then
applied both the object and hand masks.  The sky gradients were
removed by binning and fitting a plane to each image, and then
dividing the pre-processed images by this normalized sky plane.  We
then re-applied the object and hand masks, and median combined all the
frames to make the new flat field.  This process was then repeated,
except that the pre-processed data were flattened with the new flat
field.  After only a few iterations, the resulting flat field
converged.  As a further check of the robustness of our combination
technique, we made several flat field images from a random selection
of half of our sky images; pixels varied in these half-split flats
with a standard deviation of $\approx0.5\%$, consistent with the
errors expected from photon statistics (see the Appendix for a
discussion of photon statistics and other sources of photometric error
in the final images).  All pixels which were either under-illuminated
by over 50\% from the mean or which varied strongly in the half-split
flats --- these pixels were typically associated with small dust spots
or very near the edge of the CCD, and represent $<0.5\%$ of the pixels
--- were masked in the target frames.

During our processing of the sky images into a flat field, we noticed
a systematic difference between images taken during the third
observing run and those taken during the first two runs.  Frames from
the third run showed a characteristic radial dimming pattern whereby
pixels farthest away from the image center were under-illuminated by
$\approx 2\%$.  We believe this effect to be the result of a thermal
gradient, whereby the center of the CCD was cooled less efficiently,
and was therefore warmer, than the outskirts during this run.
However, we find no indication that the flat field characteristics
varied temporally within the third run.  Fortunately, because the
weather during the final run was significantly better than in the
preceding runs, a disproportionate fraction of our data was taken
during run three, giving us an ample number of exposures with which to
construct a separate flat field.  We thus generated two final flat
fields, one for the first two runs using 47 sky images, and another
for the third run from 37 sky frames.  Our target images were then
flattened by dividing by the final flat field corresponding to the run
in which they were taken.  In Section \ref{sec:finalmosaic} we show
that the final mosaic image is not affected by any run-specific
effects.

\subsection{Star Subtraction}

\begin{figure}
\plotone{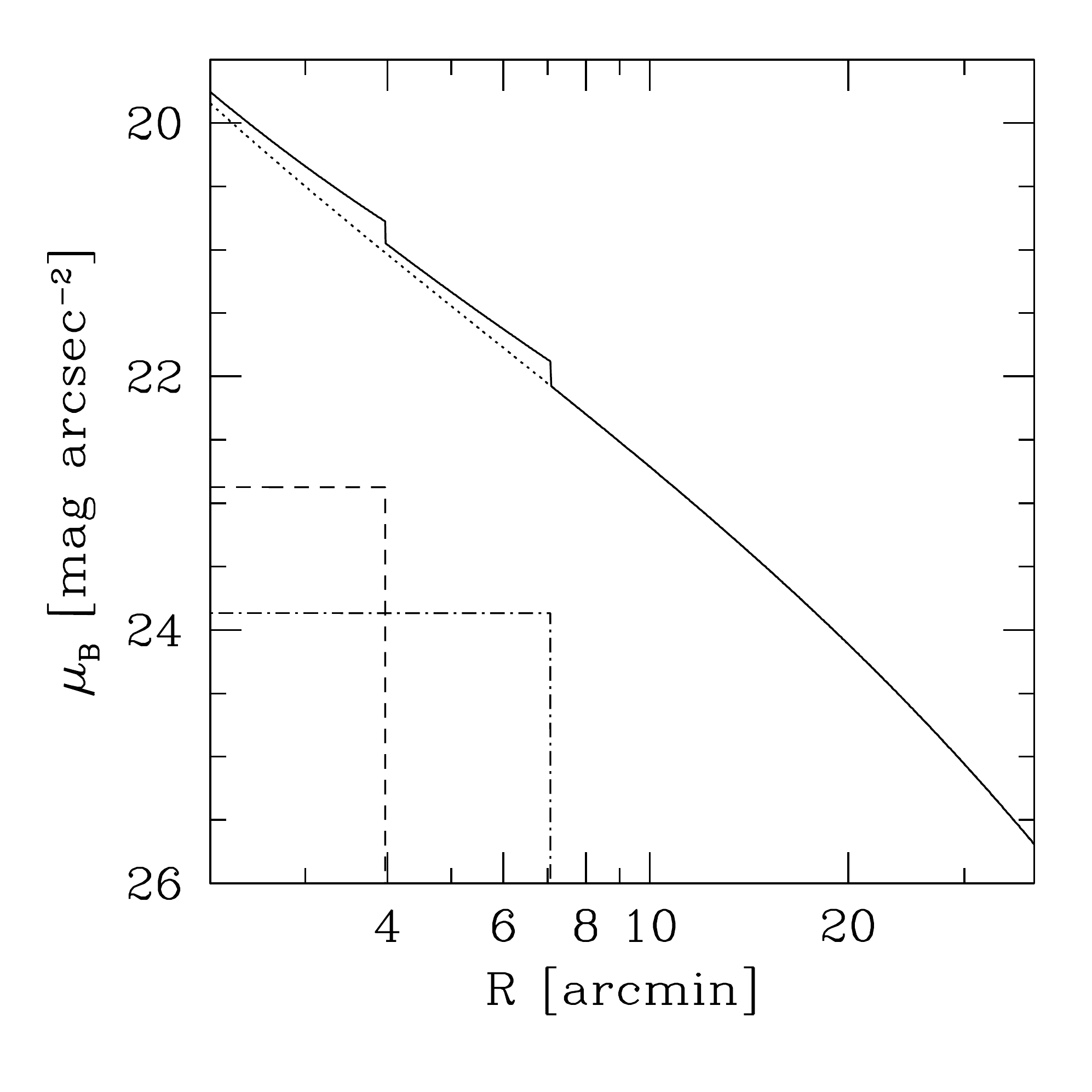}
\caption{Solid line: the extended wings of our model stellar profile
  for a $m_{B}=0$ star.  Dotted line: Our model PSF, excluding
  reflections from the dewar window surfaces.  Dashed and dot-dash
  lines: the contributions to the stellar profile from the inner and
  outer dewar window surfaces, respectively, for a star with
  axisymmetric reflections.  Unlike the data used in Slater \etal
  (2009) taken using the Washington-$M$ filter, we see no measurable
  reflections from the \Bp\ filter surfaces.
\label{fig:psf}}
\end{figure}

A key element of our data processing technique was to remove scattered
light from stellar sources in our target images using the methods
developed in Slater \etal (2009).  As Slater \etal (2009) showed, the
combination of extended stellar profiles plus scattered light from
internal reflections within the optical system can cause a spurious
diffuse signal which would contaminate the very faint ICL signal we
are trying to measure.  Moreover, the internal reflections are not
axisymmetric about the star, and are dependent on the star's position
within the field of view.  Thus, in addition to taking both our sky
and target images, during our observing runs we also took numerous
images of bright stars, varying the position of the star within the
field of view, in order to trace the stellar PSF and internal
reflections.  We acquired 13 900 s exposures of $\beta$\ Gem
($m_{B}=2.14$), five 900 s exposures of $\alpha$\ Boo ($m_{B}=1.19$),
and four 1200 s exposures of $\alpha$\ Gem ($m_{B}=2.00$), as well as
several short (1-15 s) exposures of these stars.  Using these images,
we determined the amplitude and position of reflections off the
optical surfaces and measured the stellar PSF to a radius of over
0.5\degr.  We show the extended wings of our model stellar profile in
Figure \ref{fig:psf}.

In order to subtract the stellar contribution from our target images,
for each image we created a map of the stellar light by reconstructing
the profile of each star in the image, including the
position-dependent internal reflections.  Stellar brightnesses were
measured from short (1-60 s) exposures of our target field taken
during the observing runs and the profile of each star was scaled to
match its flux.  We mapped the profile of each star out to the radius
at which it fell below 0.3 ADU and subtracted these profiles from the
individual target frames.

\subsection{Sky Subtraction} \label{sec:skysub}
In exactly the same manner as the sky frames discussed in Section
\ref{sec:flatfield}, each target image contains a sky flux level
which varies between images, and spatially across each image.
However, whereas fitting a plane to the sky signal was
straightforward for the sky images because the vast majority of their
pixels were of empty sky, the target field is permeated by large
galaxies and extended diffuse light, meaning that there is essentially
no true sky to measure.  Because the sky and diffuse light signals
vary by similar amplitudes over similar scales, it is extraordinarily
difficult to disentangle the two and remove only the sky signal.

We began the sky subtraction process by running two rounds of
\emph{objmask}, masking image artifacts such as star flares and
reflections by hand, and spatially binning the image, in the same
manner as we processed the sky images in Section \ref{sec:flatfield}.
M05 attempted to use an iterative process to minimize the
frame-to-frame scatter in the intensity of pixels at the same location
after sky subtraction.  However, subsequent testing has revealed that
our implementation of this technique achieved results not
significantly better than simply fitting a single sky plane to each
target image (see the Appendix for details).  We have therefore
utilized this simpler technique, and the sky signal was removed by
fitting a plane to each target frame in the same manner as for the sky
images.

This inability to accurately fit the night-sky signal results in our
largest source of systematic uncertainty across large scales, creating
an uncertainty of 1-2 ADU across the majority of the image, discussed
in detail in the Appendix.  Additionally, because the entire target
region is permeated with diffuse light, determining the sky zero
level, or the flux level of ``pure'' sky, was also extremely
difficult.  We selected several regions of the image which looked by
eye to be the darkest, or least diffuse light-contaminated, and
defined the sky level to be the mean pixel value in these regions.  We
emphasize that this is likely to be somewhat brighter than the true
sky level as these regions probably contain some diffuse light, and
that this could lead to a systematic over-subtraction of any
large-scale diffuse light component.  Fundamentally, these are issues
which are inherent to any surface photometry measurements in which the
objects fill the field of view, making a precise determination of the
sky flux extremely difficult.

\subsection{Creating The Final Mosaic} \label{sec:finalmosaic}
After sky subtraction, the final processing step was to simply apply a
photometric airmass correction to each target image.  The
fully-processed frames were then registered and all 102 images were
median combined into a mosaic using \emph{imcombine}, using a
3-$\sigma$ rejection to remove outlying pixel values.  All regions of
the mosaic which did not contain a minimum of five exposures were
masked.  After creating the initial mosaic, a final sky plane was fit
and removed, in the same manner as described in Section \ref{sec:skysub};
this plane had an amplitude of $\approx0.5$\ ADU across the width of
the mosaic.

As a test of the robustness of our sky subtraction and image mosaicking
procedures, we created numerous mosaics from subsets of input images
and compared these to the final mosaic.  The input frames were divided
into subsets based on various parameters, including the run, airmass,
hour angle, right ascension, declination, sky brightness, and local
time of the observation.  Only when we split the input images by right
ascension (i.e. the position of the image within the mosaic), did we
see any significant large-scale variations from the final mosaic, and
these variations were of only $\approx 0.5$ ADU.  This effect is most
likely caused by specific diffuse light features which were present in
certain areas of the mosaic which would have systematically influenced
the fit of the sky plane.

\section{\B-band Mosaic Image}

\begin{figure*}
 \plotone{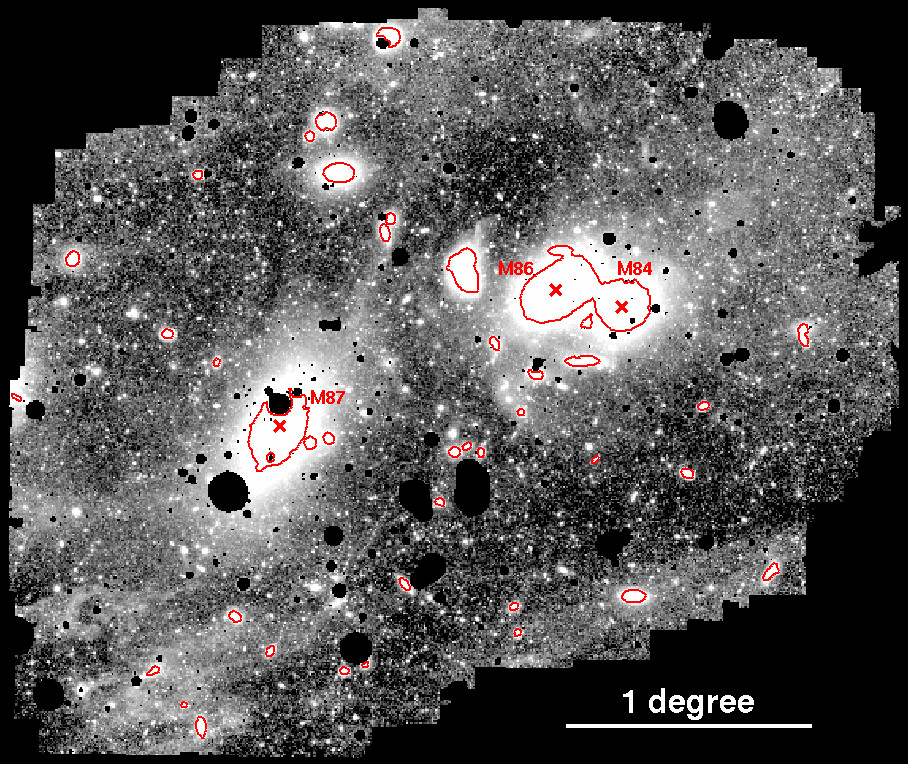}
\caption{\B-band mosaic image of the Virgo cluster core (north is up,
  east is left).  The low-surface brightness features seen in this
  image are qualitatively extremely similar to those seen in the
  \V-band imaging of M05 (see Figure 1).  All regions containing fewer
  than five exposures have been masked.  Also, stars have been masked
  out to the radius where their subtracted profile falls below 5 ADU.
  The pixels have been binned in 16\by16 groups (23.2\arcsec \by
  23.2\arcsec) in order to increase the signal-to-noise and display
  faint features.  The centers of the three giant elliptical galaxies
  in the field, M87, M86, and M84 have been marked with X's, and the
  red lines indicate the \mub=25 \magsec\ isophotes for a selection
  of the cluster's most luminous galaxies.
\label{fig:virgob}}
\end{figure*}

\subsection{ICL Features}
Figure \ref{fig:virgob} shows our final mosaic, binned in 16\by16
pixel groups (23.2\arcsec \by 23.2\arcsec), with bright stars masked.
Qualitatively, the \B-band image of the Virgo cluster core shown in
Figure \ref{fig:virgob} appears almost exactly like the \V-band image
from M05, albeit with a significantly larger mosaic area due to the
increased CCD size.  In fact, every feature identified in M05 appears
in this image, including the distorted outer halo of M87, the two
tidal streams extending northwest of M87, the doglegged plume to the
north of NGC 4435/4438, the common envelope around NGC 4413/IC 3363/IC
3349, and many other features as well (see M05 Figure 3 for a
schematic finding chart of ICL features in this field).  The
identification of these features in a second imaging survey, using a
different filter and with a modified instrumental set-up confirms the
robustness of our imaging techniques and the veracity of these
extremely faint structures.

\subsection{Galactic Cirrus} \label{sec:cirrus}

\begin{figure*}
\plotone{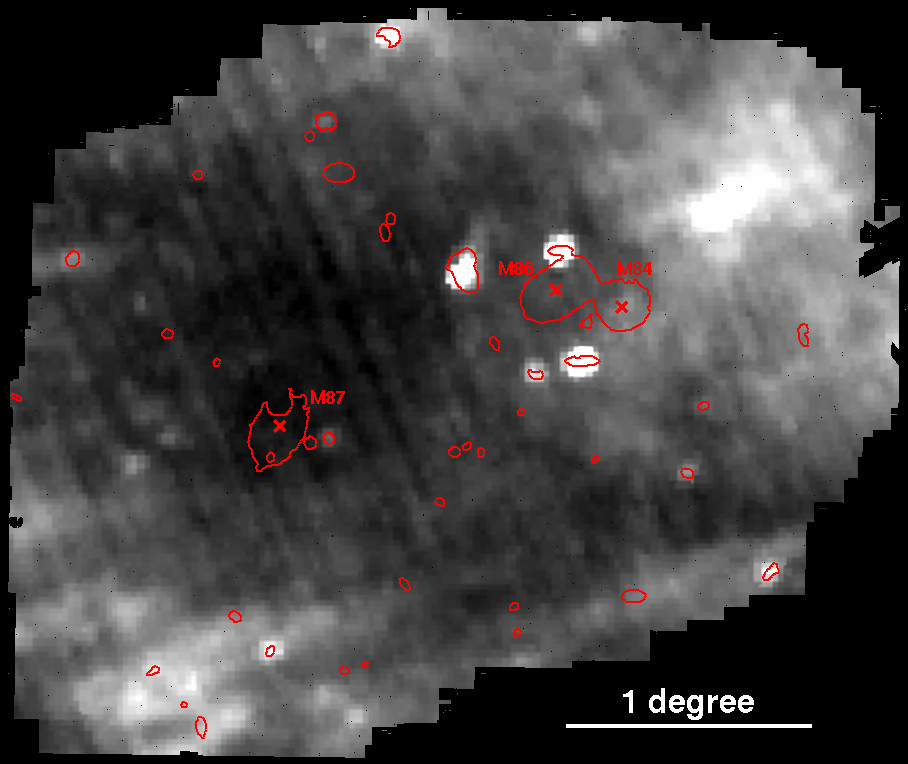}
\caption{IRAS 100$\mu$m map of the Virgo cluster core from Schlegel
  \etal (1998), on the same pixel scale as Figure \ref{fig:virgob}
  with the \B-band exposure mask applied.  As in Figure
  \ref{fig:virgob}, the centers of the three giant elliptical galaxies
  in the field, M87, M86, and M84 have been marked with X's, and the
  red lines indicate the \mub=25 \magsec\ isophotes for a selection of
  the cluster's most luminous galaxies.  Diffuse features are
  indicative of galactic cirrus, including the large knot in the
  northwest, the bright emission in the southeastern corner, and the
  long streak running nearly parallel to the southwestern border of
  the mosaic.  We find corresponding optical emission for all of these
  features, and must exclude these regions from our analysis of the
  ICL.  Fortuitously, the region to the northwest of M87 containing
  the three prominent tidal features discussed in Section
  \ref{sec:m87_streams} contains very little galactic cirrus.
  \label{fig:100mu}}
\end{figure*}

In addition to ICL in the Virgo cluster, another significant
large-scale, low-surface brightness astrophysical feature seen in this
image is galactic cirrus.  Galactic cirrus is composed of dust clouds
within the Galaxy which reflect galactic starlight (\eg Sandage 1976;
Witt \etal 2008).  These cold dust clouds are most readily detected
through their thermal emission in the far-infrared.  Figure
\ref{fig:100mu} shows the Schlegel \etal (1998) IRAS 100$\mu$m image
of our mosaic field.  The galactic cirrus features are most obviously
seen in the southeast, northwest, and southwest corners of our \B-band
mosaic and match extremely well to far-infrared emission.
Fortuitously, the core of the Virgo cluster, which contains the
majority of the ICL features we wish to measure, is relatively
unaffected by galactic cirrus.  However, the cluster core is
surrounded by a ring of cirrus which obscures any ICL which may lie
behind it.  These galactic cirrus features were not as readily
apparent in the \V-band image of M05 since most of the cirrus lies
outside of the smaller field of view of that data set.  Because these
cirrus features have no consistent optical color (Guhathakurta \&
Tyson 1989; Witt \etal 2008) and vary in intensity on very small
scales, we do not attempt to model and subtract them from the data and
must simply ignore any areas affected by these features for the
purposes of measuring ICL.

\section{ICL \BV\ Colors}\label{sec:color}

\begin{figure*}
\plotone{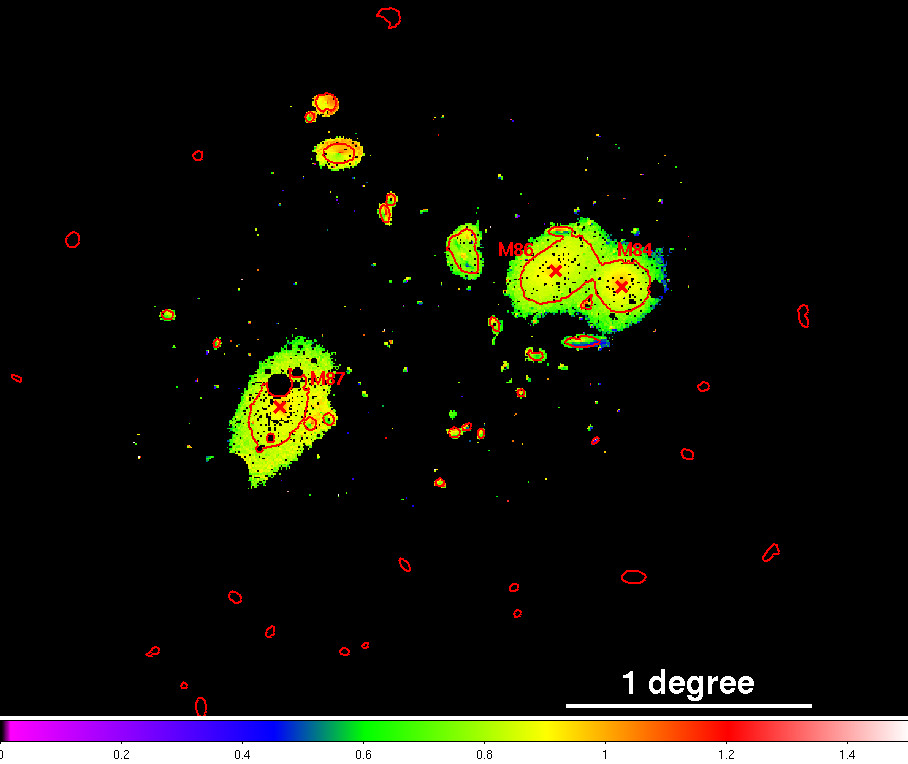}
\caption{\BV\ color map of the Virgo core in the regions of overlap
  between our two mosaic images, on the same pixel scale as Figure
  \ref{fig:virgob}.  As in Figure \ref{fig:virgob}, the centers of the
  three giant elliptical galaxies in the field, M87, M86, and M84, have
  been marked with X's, and the red lines indicate the
  \mub=25 \magsec\ isophotes for a selection of the cluster's most
  luminous galaxies.  Bright stars and all pixels with \muv$>27.0$
  \magsec\ or \mub$>27.5$ \magsec\ have been masked.  As discussed in
  detail in the text, features at lower surface brightness are subject
  to large systematic errors which cause measurements of their colors
  to be unreliable over large scales.  The colors in the central
  regions of galaxies have an asymmetric profile due to a small
  ($\simlt1$\arcsec) coordinate registration offset between the \B\
  and \V\ images.
\label{fig:color}}
\end{figure*}

By combining our \V-band imaging data from M05 with our current
\B-band dataset, we can measure the \BV\ colors of the diffuse light
in the Virgo cluster core.  However, because we have estimated the sky
emission differently for our \B-band data than for the \V-band data
presented in M05 (described in Section \ref{sec:skysub}), for consistency
we have re-reduced that data using the new sky subtraction method.
This has only a very minor effect on the resulting mosaic and all the
features described in that work remain qualitatively similar; a more
quantitative discussion of the differences between these two sky
subtraction techniques can be found in the Appendix.

A major challenge in generating a large-scale color map of the cluster
is setting the proper sky flux level for both images.  Not only is the
absolute image sky level difficult to determine precisely (see Section
\ref{sec:skysub}), but within each image we also have large-scale sky
level uncertainties which we estimate to be of order 1-2 ADU (see the
Appendix).  Because the features we are measuring have very low flux
levels, even small errors in the sky levels can translate into
substantial differences in the final colors.  For example, at \muv=28
\magsec, a sky level offset of just 1 ADU becomes a 0.4
mag. uncertainty in the \BV\ color.  At higher flux levels this effect
is significantly reduced as the sky level uncertainty becomes a very
small fraction of the total flux; at \muv=26 \magsec, a 1 ADU zero
level offset translates to only a 0.07 mag. uncertainty in the \BV\
color.  Figure \ref{fig:color} shows our \BV\ color map in the region
of overlap between the two mosaic images, which is limited primarily
by the \V-band areal coverage.  The image is shown on the same pixel
scale as Figure \ref{fig:virgob}, with all pixels at surface
brightness \muv$>27.0$ \magsec\ or \mub$>27.5$ \magsec\ masked.

Each of the three giant elliptical galaxies --- M87, M86, and M84 ---
shows a very red central core with a distinct radial gradient trending
bluer in the outskirts, in excellent agreement with previous studies
(\eg Carter \& Dixon 1978; Davis \etal 1985; Goudfrooij \etal 1994;
Bernardi \etal 2003; Liu \etal 2005).  In this image, however, each of
these and several other galaxies shows an azimuthally varying color
gradient in the innermost regions.  This is an instrumental effect
caused by very small (sub-arcsecond) offsets in the coordinate
registrations of the \B\ and \V\ images, coupled with the very steep
luminosity gradients present in the galactic centers.  The two data
sets were taken several years apart, with different CCDs, on a
telescope which underwent substantial modification to its optical
system in the intervening years.  The resulting high-order distortions
make it exceedingly difficult to register the entire field of view to
the sub-arcsecond precision needed to precisely measure the rapidly
varying inner regions of galaxies, and our survey was never designed
to accomplish this.  Instead, our analyses are concentrated on the
diffuse outer regions of galaxies which are not effected by these
minute registration offsets, due to the large angular sizes of these
features.

At very low surface brightnesses the sky level uncertainties discussed
above make any determination of the color from such a large-scale map
highly problematic.  Indications of these effects can indeed be seen
in the upper left corner of this image, particularly in the galaxies
NGC 4473 and 4477, which appear particularly red.  Due to the specific
geometry of the images which compose our mosaic, this is the region in
which our sky level uncertainties are largest.  We believe our sky
level uncertainties in this area to be up to 3-4 ADU, which could
easily lead to uncertainties of several tenths of a magnitude in \BV\
color, even at the relatively high surface brightnesses shown.  At
surface brightnesses fainter than the \muv$>27.0$ \magsec, \mub$>27.5$
\magsec\ limits shown in Figure \ref{fig:color}, these issues become
even more pronounced, making such a large scale map an unreliable
source for determining the color of ICL features.

In order to better measure the colors of interesting ICL features, we
have developed two methods with which to measure colors and estimate
the photometric errors, depending on the features' angular scales.
For large, degree-scale objects, such as the extended stellar envelope
of M87, the dominant uncertainty comes from our sky level gradients,
which we can estimate and include in our error model.  Over the
smaller, $\simlt10$\arcmin\ scales of individual tidal streams, other
systematic effects dominate, and we have developed a local background
subtraction technique which allows us to more robustly estimate and
reduce systematic uncertainties.  The sections below provide examples
of these techniques and measure the colors of many of the image's most
interesting low-surface brightness features, while the error budgets
are described in the Appendix.

\subsection{Extended Stellar Envelope of M87}\label{sec:m87profile}
Our analysis of the luminosities, shapes, other \V-band photometric
properties of M87 and other Virgo giant ellipticals can be found in
Janowiecki \etal (2010).  Here, we focus solely on the \BV\ color
properties of M87.  Because the outer regions of the other two giant
elliptical galaxies in the field, M84 and M86, overlap one another in
projection, we have not attempted similar analyses for these galaxies.

\begin{figure}
\plotone{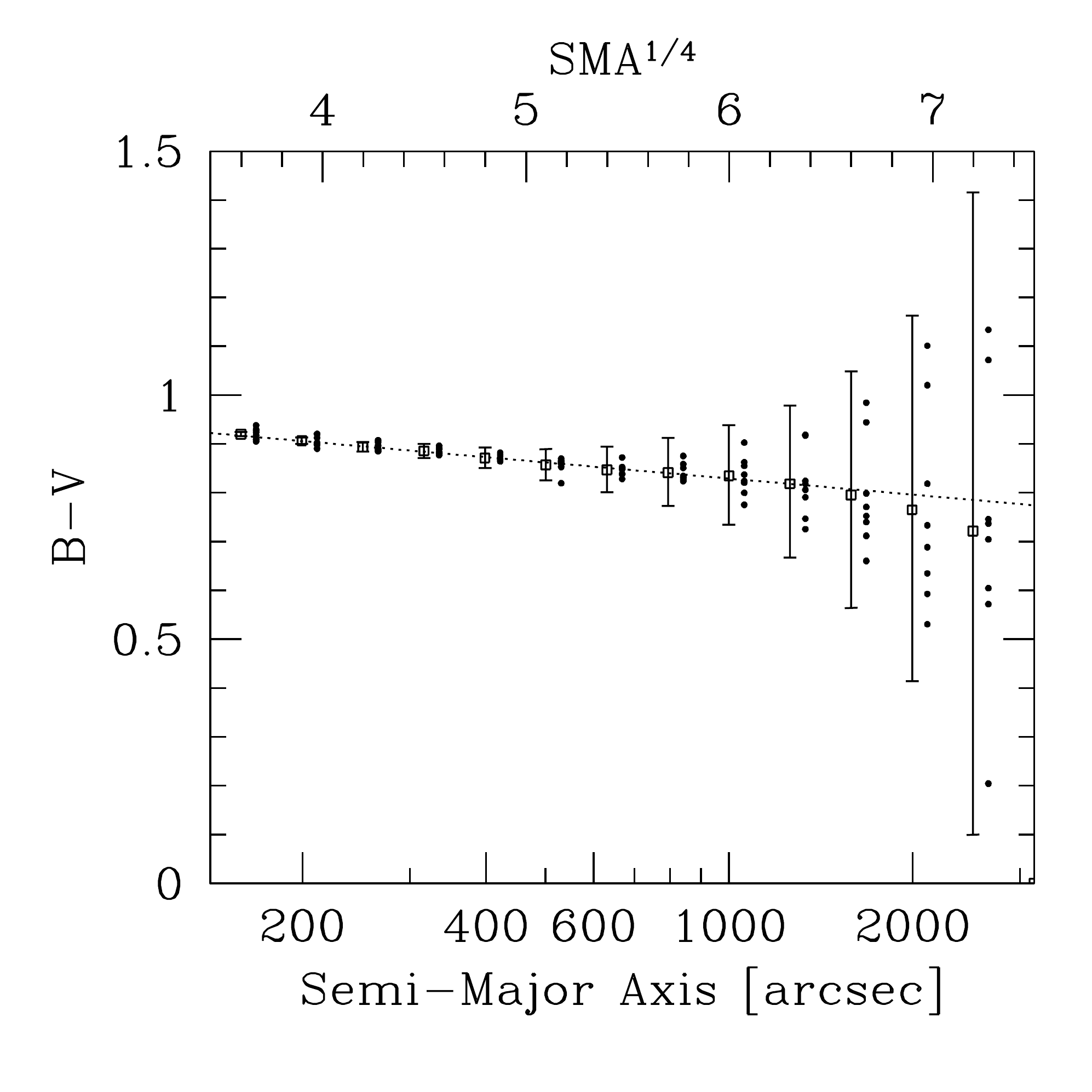}
\caption{Color profile of M87's extended stellar halo.  Squares are
  the color using all pixels in the bin, with $\pm1$ ADU error bars.
  Circles show the azimuthal color variance by measuring the color in
  eight equal angle slices at each radius (see text for details),
  offset slightly right for clarity.  The dotted line is a fit to the
  color profile inside of 1000\arcsec, and has a slope of $\Delta(B-V)
  / \Delta(\log(SMA))=-0.11$.
\label{fig:m87profile}}
\end{figure}

We have measured M87's color in elliptical annular bins, defined by
the galaxy's \V-band isophotes as measured by Janowiecki \etal (2010).
As in that work, before measuring the galaxy luminosity, small-scale
sources were masked using \emph{objmask}.  To effectively detect these
sources, we subtracted a ring-median smoothed image from the original,
thus removing the large-scale galaxy luminosity distribution and
leaving only the small-scale sources (see Janowiecki \etal 2010 for
details).  All sources detected in either image were masked.  Figure
\ref{fig:m87profile} shows our resulting color profile.  The open
squares show the mean color within each annulus.  The photometric
errors in the color measurements are entirely dominated by the
systematic uncertainty in the sky level calibration, which we estimate
to be $\pm1$ ADU on these scales (see the Appendix); the error bars in
Figure \ref{fig:m87profile} reflect this $\pm1$ ADU uncertainty.
Consistent with previous studies, our results show a bluing trend with
radius.  The dotted line in Figure \ref{fig:m87profile} shows a fit of
the color profile inside of 1000 arcsec, which has a slope of -0.11 in
$\Delta(B-V) / \Delta \log(SMA)$.  Our measured color profile of M87's
outer regions is in excellent agreement with the results of Liu \etal
(2005), and is consistent with a continuation of the color profile of
the galaxy's inner regions measured by Zeilinger \etal (1993).

To measure the azimuthal variance in the color, we have divided each
elliptical annulus into eight equal angle slices, with the color
results plotted as solid circles in Figure \ref{fig:m87profile}.  At
all radii beyond 400\arcsec, the measured colors in all the angular
slices lie within our $\pm1$ ADU systematic error.  Inside of this
radius, the azimuthal variance is dominated by the coordinate
registration offsets discussed above.  A closer examination of the
measured colors in the outermost, lowest-surface brightness regions of
the galaxy reveals that the color varies smoothly around the galaxy,
indicating that the azimuthal variance is caused by a systematic
gradient, such as that produced by sky level uncertainties, as opposed
to random fluctuations in the photometry.  Complicating these
measurements even further is the presence of galactic cirrus which
overlaps with M87's outermost reaches to the south and east.  As
described in Section \ref{sec:cirrus}, it is extremely difficult to
separate the two components, and any photometric measurements of this
area may, in fact, be dominated by the cirrus over M87's stellar halo.
Removing the data from the most cirrus-contaminated areas, however,
does not significantly improve the color scatter of the angular slices
in Figure \ref{fig:m87profile}.

\subsection{Colors of Streams}\label{sec:streamphot}

\begin{figure*}
\plotone{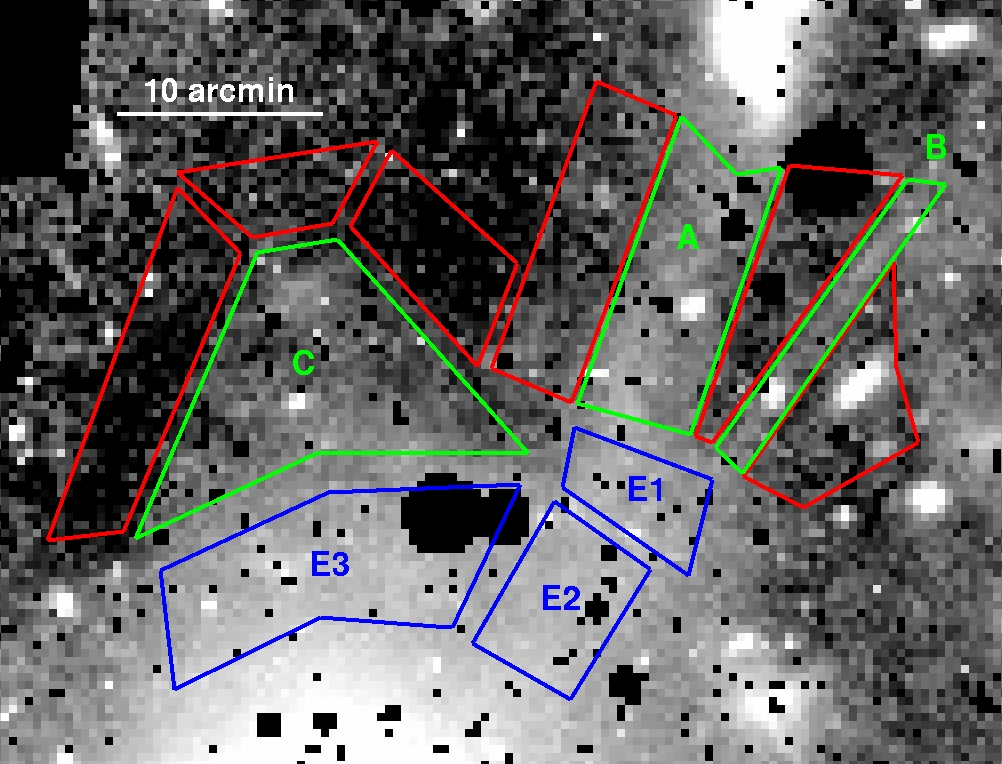}
\caption{The area just to the north of M87, containing several
  interesting tidal features for which we have measured \BV\ colors.
  The regions outlined in green -- labeled A, B, and C, respectively
  --- are the tidal features themselves.  The blue regions --- labeled
  E1, E2, and E3 --- are regions within M87's stellar envelope.  The
  red regions are examples of background regions used to measure the
  local sky brightness.  Discrete objects in the field, such as bright
  galaxies, are masked during the photometric analysis.  Details of
  the photometric techniques used on these regions are given in the
  text.
\label{fig:m87_streams}}
\end{figure*}
\begin{figure}
\plotone{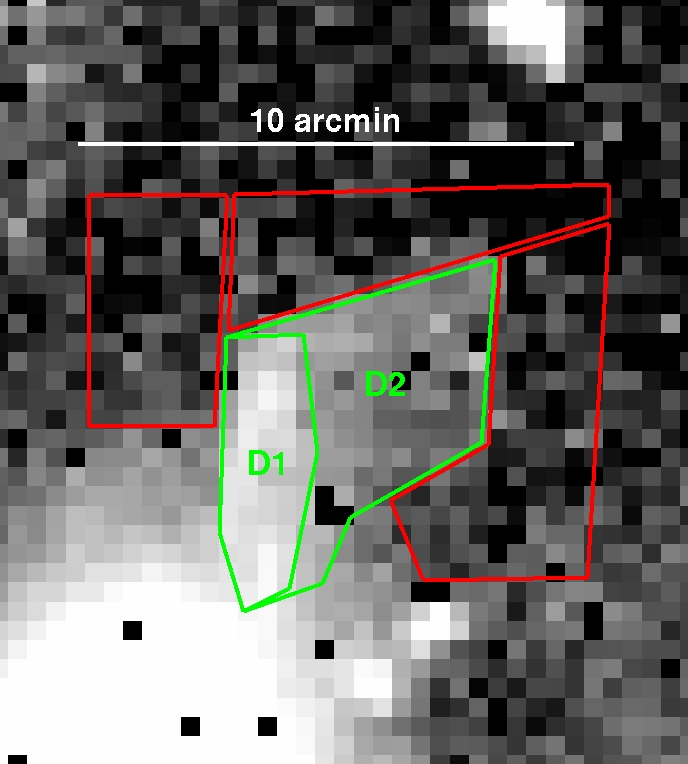}
\caption{The plume to the north of the interacting galaxy pair
  NGC4435/4438.  Similar to Figure \ref{fig:m87_streams}, the green areas
  --- labeled D1 and D2 --- enclose the plume itself, while the red
  regions are used as the local background regions.
\label{fig:ngc4435_streams}}
\end{figure}

While our ability to accurately measure colors is limited over large
scales by gradients in the photometric sky levels and extended diffuse
light sources, over shorter scales these uncertainties are much
smaller.  Thus, for smaller scale features, such as individual ICL
streams which span $\simlt10$\arcmin, we can use regions adjacent to
the features to define the local background flux and more precisely
measure luminosities and colors.  We have used a differential
photometry technique whereby we define \emph{object} and neighboring
\emph{background} regions, and measure the object luminosity by
subtracting the flux in these local background fields.  Discrete
objects in the fields, such as bright galaxies, are masked so that we
are measuring only the diffuse light of the features themselves, and
we vary a number of parameters, including the precise placement of
region boundaries, in order to estimate the measurement uncertainties
(these processes are described in detail in the Appendix).  In
essence, this technique is a variation on familiar aperture photometry
methods used for point sources, where the background flux is measured
from an annulus just beyond a circular aperture.

The background regions contain three sources of flux, each of which is
also present in the object region and should be removed to isolate the
stream's luminosity.  First, as discussed previously, the background
regions should share any systematic offset in the sky zero level with
the object.  Second, in addition to individual tidal streams, the
cluster ICL likely contains a large-scale diffuse component which
permeates both the object and background regions.  Finally, there are
numerous foreground and background sources in all the regions;
although we attempt to mask these objects (described in the Appendix),
because faint sources should be distributed approximately
homogeneously on these scales (Brainerd \etal 1995; Villumsen \etal
1997; Connolly \etal 2002; Coil \etal 2004; Morganson \& Blandford
2009), their contribution should be equivalent in all regions.  We
discuss a detailed error budget for this photometric technique in the
Appendix.

\subsubsection{M87's Northern Stellar Envelope and Tidal Streams} \label{sec:m87_streams}
A number of the most interesting low-surface brightness features in
our mosaic occur in the northernmost regions of M87's extended stellar
halo.  In particular, Figure \ref{fig:m87_streams} highlights three
ICL features that we have identified for more detailed study: the
large tidal stream extending north-west from M87 toward NGC 4461/4458
(stream A from M05); the smaller stream just to the west which is also
emanating from M87 to the northwest (stream B from M05); and the broad
plume to the east of these streams, almost directly north of M87's
core (unlabeled in M05, labeled C in Figure \ref{fig:m87_streams}).
Additionally, we have defined three neighboring regions (labeled E1,
E2, and E3) which are more clearly part of M87's outer stellar
envelope.  In addition to being discussed below, all color
measurements for these features can be found in Table
\ref{tab:streamcolors}.

\begin{deluxetable}{lcc}
  \tablecolumns{3}
  \tablewidth{0pc}
  \tablecaption{Photometry of Low-Surface Brightness Features
   \label{tab:streamcolors}}
  \tablehead{
    \colhead{Region}                          & \colhead{$B-V$} & \colhead{$\mu_{V}$} \\
    \colhead{(see Figures \ref{fig:m87_streams} \& \ref{fig:ngc4435_streams})} & \colhead{}      & \colhead{mag arcsec$^{-2}$}
  }
  \startdata
    A  & 0.75-1.05 & 28.6 \\
    B  & 0.8-1.2   & 29.2 \\
    C  & 0.7-1.0   & 28.7 \\
    E1 & 0.8-0.9   & 28.2 \\
    E2 & 0.75-0.85 & 27.7 \\
    E3 & 0.75-0.85 & 27.6 \\
    D1 & 0.65-0.75 & 27.2 \\
    D2 & 0.5-0.7   & 28.4 \\
  \enddata
\end{deluxetable}

Precisely measuring the colors of these features is quite difficult,
and the uncertainties are dominated by numerous systematic effects
which are detailed in the Appendix.  For this reason, we prefer to
quote our measured colors as ranges, rather than as central values
with error estimates, which imply a well-characterized Gaussian error
model.  For features A, B, and C, we measure \bv\ colors of
$0.75-1.05$, $0.8-1.2$, and $0.7-1.0$, respectively, with mean surface
brightnesses of \muv$\approx28.6$ \magsec, \muv$\approx29.2$ \magsec,
\muv$\approx28.7$ \magsec, respectively.  If these streams were
displayed on the raw color map of Figure \ref{fig:color}, however,
they would each show a \bv\ color of $\approx0.5$.  This difference in
color highlights the difficulty in interpreting the colors of very low
surface brightness features from the raw \BV\ color map and the
importance of using a local background subtraction when performing
photometric measurements.

Regions E1, E2, and E3 in M87's stellar envelope display very similar
colors to the tidal features.  For E1 we measure \bv$\approx0.8-0.9$
with mean surface brightness of \muv$\approx28.2$, while both E2 and
E3 have \bv$\approx0.75-0.85$, with surface brightnesses of
\muv$\approx27.7$ \magsec and \muv$\approx27.6$ \magsec, respectively.
The measured color ranges are tighter in these regions due to their
higher surface brightnesses, meaning that systematic effects have a
comparatively smaller effect (see the Appendix for details).  Both E2
and E3 lie approximately along the $2000\arcsec$ semi-major axis
ellipse around M87, and their colors measured using this differential
method agree very well with those of the radial profile, shown in
Figure \ref{fig:m87profile}, at this radius.

Thus, within the measurement errors, we find that the optical colors
of the tidal features are consistent with those of M87's outer stellar
envelope.  This suggests that these various features consist of
similar stellar populations and may have a common origin.  In fact,
the colors we measure in these features are also consistent with the
optical colors measured for Virgo's dwarf elliptical galaxy population
of \bv$\approx0.8$ (van Zee \etal 2004).  The implications of these
findings for the evolution of galaxies within the Virgo cluster are
discussed in Section \ref{sec:discussion}.

\subsubsection{The Plume of NGC 4435/4438}
Figure \ref{fig:ngc4435_streams} shows another prominent low-surface
brightness feature in the Virgo cluster core, the dog-legged plume
just to the north of the interacting pair NGC 4435/4438 (feature D
from M05).  For this plume, we have divided the feature into two
sections, the relatively high surface brightness vertical plume, and
the lower surface brightness diffuse emission to the west, labeled D1
and D2, respectively, in Figure \ref{fig:ngc4435_streams}.  These
features are the bluest in the image for which we have made reliable
measurements.  We find that D1 has a \BV\ color of $\approx 0.65-0.75$
with mean surface brightness \muv$\approx27.1$ \magsec\ while D2 has
\BV$\approx0.5-0.7$ and \muv$\approx28.4$ \magsec.

A wide range of multi-wavelength observations have recently been made
of this feature (Cortese \etal 2010; Krick \etal 2010), and its origin
--- whether a tidal feature from the interaction of NGC 4435/4438 or
simply foreground galactic cirrus --- is currently in doubt.  While
its visual morphology is immediately reminiscent of an
interaction-induced tidal plume, especially given that NGC 4438
displays obvious signs of tidal disturbance, Cortese \etal (2010)
argue that the feature's UV-IR color and extremely narrow CO and HI
velocity widths are more consistent with it being a galactic cirrus
dust cloud.  If it is truly an extragalactic feature, its blue colors
may hint at recent tidally induced star formation, and HST imaging
should be able to resolve a young stellar population.  If it proves to
be a galactic cirrus feature, however, it provides an excellent, if
sobering, example of the difficulty of discerning an object's origin
by morphology alone and of the problems that galactic cirrus poses for
deep observations of extragalactic objects.

\subsubsection{Limits of the Differential Photometry Technique}
While our differential photometry technique allows us to increase the
precision of our color measurements by more robustly estimating the
photometric uncertainties, the procedure's applicability is often
limited by features' geometry.  Most importantly, any features on
which we use this technique must have obvious background regions
immediately surrounding them, preferably on multiple sides.  As
detailed in the Appendix, the choice of background region is often the
dominant source of systematic uncertainty in a feature's measured
color.  This geometric criterion rules out many of the most
interesting ICL features seen in our image as candidates for the
differential photometry method.  For instance, a region of our mosaic
displaying a number of interesting low-surface brightness tidal
features is the area surrounding M86 and M84, especially the many
interacting galaxies to their south. However, because there are
numerous overlapping ICL features throughout the region, including
both individual tidal streams as well as larger diffuse components, it
is extraordinarily difficult to properly define regions in which to
measure the background flux for any particular feature, thus
limiting our ability to perform accurate photometry of these features.

\subsection{Fundamental Photometric Limits}
The extremely large color scatter in the azimuthally sliced bins of
the outermost regions of M87's extended stellar halo seen in Figure
\ref{fig:m87profile} suggests that we are unable to measure the colors
of diffuse light at these very low surface brightnesses with the
precision necessary to meaningfully constrain the stellar populations
present.  We estimate the surface brightness limit to which we can
adequately measure colors to be \muv$\approx27$ \magsec\ in large,
degree-scale features such as the giant elliptical galaxies.  While
our local background subtraction technique detailed in Section
\ref{sec:streamphot} is able to substantially reduce the photometric
errors for certain features with smaller angular sizes ($\simlt10$
arcmin), allowing us to push our measurements to lower surface
brightnesses, it is not universally applicable to all diffuse light
features in the image.  Because at the lowest surface brightnesses we
are limited by systematic uncertainties in the brightness of both the
night sky and astrophysical sources of diffuse light, and not from
random statistical noise or known features of the optical design, we
may be approaching the fundamental limit of precision available from
ground-based, wide-field surface photometry.

While the Virgo cluster is an appealing target in which to study ICL
in part because it is the nearest massive galaxy cluster and thus ICL
features can be seen over large angular scales, its large angular size
actually acts as a hindrance to our ability to perform accurate
surface photometry at the faintest levels.  As discussed in Section
\ref{sec:skysub}, even with the Burrell Schmidt's extremely wide-field
imaging capability, the diffuse light features of the Virgo cluster
essentially fill the entire field of view, making it extremely
difficult to measure and subtract the brightness of the night sky.
Thus, we may be able to achieve even greater imaging depth and reduce
large-scale uncertainties for targets with smaller angular sizes, such
as individual galaxies in the local universe or clusters at higher
redshift, where the object does not fill the field of view, and the
sky brightness, including large-scale sky gradients, can be more
precisely measured.

\section{Summary and Discussion} \label{sec:discussion}
In this paper, we have presented our \B-band image of the Virgo
cluster core, taken as part of our ongoing deep imaging survey of the
Virgo cluster.  All aspects of the survey, from the optical and
mechanical design of the telescope system, to the acquisition,
analysis, and reduction of the data have been optimized to detect
diffuse light at low surface brightness.  Our final photometric
uncertainties are dominated by systematic errors resulting from sky
subtraction and astrophysical backgrounds and not simple photon
statistics.

Our \B-band imaging confirms the results of M05, which found a vast
web of diffuse light features in the core of the Virgo cluster.  These
features include large-scale diffuse components, as well as a number
of discrete features such as streamers, arcs, tails, and plumes which
provide a wealth of information on the dynamical history of the
cluster and its constituent galaxies.  Just outside the cluster core,
however, we have detected a large number of galactic cirrus features
which prevent us from studying diffuse extragalactic sources which lie
behind them.

By combining these \B-band results with the \V-band imaging from M05,
we have been able to measure the colors of a number of low surface
brightness features in Virgo's core.  We have measured the radial
color profile of M87 out to very large radius, demonstrating that the
color gradients seen in the inner regions of the galaxy extend to its
outer stellar envelope.  Furthermore, we have measured the colors of
several tidal features which extend from M87's stellar envelope and
find that the two populations have similar optical colors within the
measurement uncertainties.  These results are consistent with the
majority of other measurements of ICL color, which suggest that the
ICL colors should be similar to those of the outskirts of the
cluster's brightest galaxies (\eg Zibetti \etal 2005; Sommer-Larsen
\etal 2005; Krick \& Bernstein 2007).

These results are consistent with the hypothesis that the outer
envelopes of cD galaxies like M87 may be predominantly built-up by the
tidal stripping and disruption of smaller satellite galaxies, the same
mechanism which likely generates the tidal streams and other ICL
features (Rudick \etal 2009).  In this scenario, the cD envelope is
simply composed of tidal streams of ICL which have been mixed in the
cluster potential, and dissolved to form a smooth distribution.  While
optical colors alone cannot definitively prove that any two stellar
populations are identical due to the well-known age-metallicity
degeneracy, the similar colors of these two populations is suggestive
of a common origin.  

The optical colors of both the tidal features and M87's outer envelope
are also consistent with the colors of the Virgo cluster's dwarf
elliptical galaxy population (van Zee \etal 2004).  Thus, the
disruption of these galaxies in the cluster potential may provide a
ready source for the intracluster stars which make up the low-surface
brightness outer envelope of M87 and the tidal streams.  In fact,
the thinness of two of the tidal streams studied in Section
\ref{sec:m87_streams} suggests that they originate from low-mass
galaxies with small velocity dispersions (M05), such as dwarf
ellipticals.  Furthermore, the results of Williams \etal (2007), who
measured the age and metallicity of Virgo's intracluster stars at much
larger radius from any of the cluster's large galaxies to be $\simgt
10$ Gyr and [M/H]$\simlt -1.0$, are consistent with the same
\bv$\approx 0.8$ color that we measure for the streams and stellar
envelope immediately beyond M87 (Bruzual \& Charlot 2003).  This color
similarity further bolsters the case that M87's extended stellar
envelope, the tidal streams from disrupted galaxies, and the truly
intra-cluster stars all form a single population, formed through
similar mechanisms of tidal stripping during the cluster's
hierarchical assembly.

\acknowledgements The authors greatly appreciate Xu Zhou and Jiansheng
Chen for sharing their electronic data of M87's surface brightness
profile from Liu \etal (2005) with us.  We thank Adolf Witt for
helpful discussions on the properties of galactic cirrus.  We also
thank Luca Cortese for providing an early draft of his manuscript on
the NGC4435/4438 feature, and Jessica Krick for discussions on the
nature and extent of this feature. This paper benefited from the
comments of an anonymous referee, especially in the presentation of
the maps in Figures 2-4.  CSR appreciates support from the Jason
J. Nassau Graduate Fellowship Fund and the Sigma Xi GIAR program.  JCM
is supported by the NSF from grants AST-0607526 and AST-0707793.

{\it Facility:} \facility{CWRU:Schmidt}

\appendix

\section{Photometric Errors}
There are a number of both random and systematic effects which
contribute to the uncertainty in our photometric measurements, many of
which were mentioned in the main body of the paper.  Here, we discuss
how these various effects interact to yield our final measurement
errors.  In general, we find that random errors from photon noise make
an insignificant contribution to the error budget, and that the
systematic effects which dominate are highly dependent on the angular
scales being probed.  Below, we calculate the errors from photon
statistics to show that they are negligible, and then describe the
error sources which predominate in the angular scale regimes
discussed in Sections \ref{sec:m87profile} and \ref{sec:streamphot}.

\subsection{Small-scale Random Errors}
The pixel-to-pixel measurement error based simply on random photon
statistics is a straightforward calculation.  For a single image with
a typical sky value of 750 ADU (because the ICL features we are
measuring are $<1\%$ of the sky brightness, we are dominated by sky
photons even in our target frames), the photon noise in ADU is given
by $\sqrt{C_s/g}$, where $C_s$ is the sky flux in ADU and $g$ is the
gain in $e^{-}$ ADU$^{-1}$.  For our gain of 2.0$e^{-}$ ADU$^{-1}$
this comes to 19.4 ADU \ppix, or $2.6 \%$.  When we combine images,
however, this error is scaled by a factor of $1.22/\sqrt{N}$, where
$N$ is the number of images combined.  Our flat field, which contains
a minimum of 37 images, thus has an error of $0.52\%$, corresponding
to 3.9 ADU \ppix.  In the final mosaic, we further reduce the errors
by combining a minimum of five images, for a final error of $0.28\%$,
or 2.1 ADU \ppix.  This is, however, an absolute maximum for our
random photometric error, and for all measurements made in this paper
it is significantly lower.

In the calculation above we assumed the minimum number of five target
exposures, whereas $\simgt30$ exposures are more typical across the
image, which reduces the error by another factor of 2.6, or to $<1$
ADU \ppix.  Even greater reductions in the random errors, however, are
achieved due to the fact that the diffuse nature of ICL features means
that they are spread over a very large number of pixels.  Any feature
which covers even 100 pixels will have its random photometric error
reduced by an order of magnitude.  Because the final mosaic shown in
Figure \ref{fig:virgob} has been binned in 16$\times$16 pixel groups,
its random per pixel photometric error from photon statistics falls to
well below to 0.1 ADU, or $<0.01\%$.

\subsection{Large-scale ($\simgt0.5$\degr) Systematic Sky Level
  Uncertainties} 

\begin{figure*}
\plotone{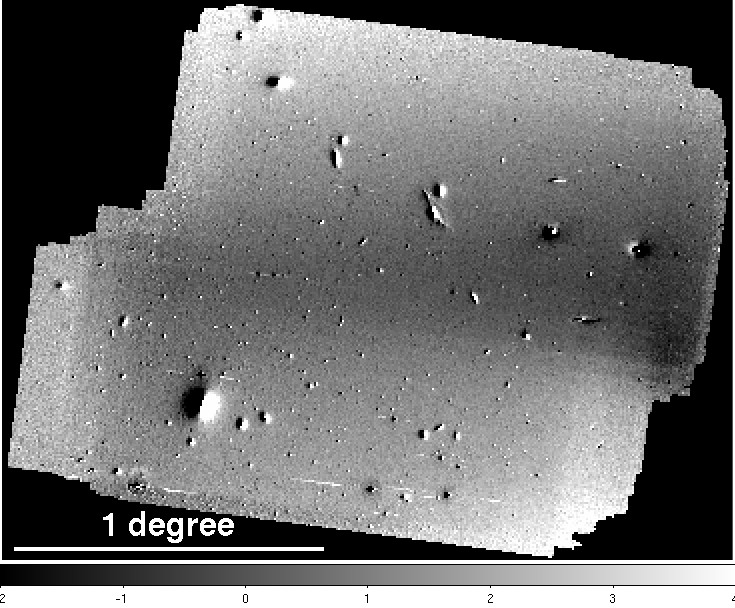}
\caption{The difference between the current image using a new sky
  subtraction technique and the 2005 image, in ADU.  The butterfly
  pattern seen near the centers of bright galaxies is caused by a
  slight differences in the astrometric calibration of the two images
  (see Section \ref{sec:color}).
\label{fig:skydif}}
\end{figure*}

As mentioned in Section \ref{sec:skysub}, our ability to properly subtract
the flux of the night sky from our object images leads to our greatest
source of large-scale systematic uncertainty.  In that section, we
described the two different techniques that we have used to measure
the sky signal.  Figure \ref{fig:skydif} shows the difference between
our \V-band images, sky subtracted using the two methods.  We expect
the large-scale sky level uncertainties in the \V-band mosaic to be
more severe than in the \B-band, since the \V-band data was taken
using a CCD with only half of the current field of view, meaning that
there were even fewer pixels available from which to estimate the sky
level gradients.  In general, the sky level offset varies smoothly
across the field.  The largest differences are found in the extreme
northeast and southwest corners, where the two images differ by up to
4 ADU, but values of $\approx1-2$ ADU are much more common.  While in
certain specific cases, we see that one of the techniques yields
unphysical results in certain areas of the image (\eg the extremely
red colors of NGC 4473 and 4477 seen in the northeast corner of
Figure \ref{fig:color}), we cannot constrain the sky level more
generally.  Thus, when measuring large-scale features, such as the
extended stellar envelope of M87 in Section \ref{sec:m87profile}, we can
only roughly estimate a systematic sky level uncertainty, using the
variations seen in Figure \ref{fig:skydif} as a guide, and recognizing
that this is not the only source of systematic uncertainty.

\subsection{Intermediate-scale ($\simlt10$\arcmin) Differential
  Photometry Uncertainties} 

In Section \ref{sec:streamphot} we described
the differential photometry technique which we have developed to
measure the colors of low-surface brightness features on intermediate
scales, such as individual tidal features.  Using this method, we can
more robustly measure the colors of diffuse light features, resulting
in much smaller uncertainties.  We have identified four primary
contributions to the uncertainty using this technique, each of which
behaves in systematic, highly non-Gaussian ways.  While the precise
magnitude of these effects varies depending on the feature being
measured, we use stream A, shown in Figure \ref{fig:m87_streams}, as
an example to illustrate the relative importance of each.  For this
feature, we find a color of \BV$\approx0.75-1.05$, which might be
re-written as \BV$=0.9 \pm 0.15$.

In order to measure a feature's color, we need to determine the per
pixel flux level of both the object and background regions.  These
measurements are limited by the uniformity of flux across the regions
from astrophysical sources.  While in the idealized case of a Gaussian
distribution the precision to which we could determine the mean value
is given by $\sigma / \sqrt{N}$, which for our data comes to
$\approx0.01$ ADU, the non-Gaussianity of the pixel-values results in
a much larger error.  We have estimated this uncertainty by using
various statistical measures of the per pixel flux, including the
assigning the per pixel flux to be the mean or median value.  In
general, we find that our results are relatively insensitive to the
statistical measurement used, and for our example stream this results
in a \BV\ error of $\pm \approx 0.05$ mag.

There are two distinct effects contributing to the photometric
uncertainty related to the placement of the object and background
regions, respectively.  The region boundaries were determined by eye,
and are designed to qualitatively select the stream itself and the
surrounding background regions where there are no obvious ICL
features.  In general, defining the object regions was relatively
straightforward, and we find that almost any reasonable selection for
the regions' boundaries have a comparatively small effect on the
measurement uncertainty, which we estimate to be $\pm\approx0.05$ mag
in the \BV\ color for our example stream.  The uncertainty resulting
from selecting various background regions, however, has a profound
effect on the measured colors, and is often the single largest source
of error.  Each background region has a slightly different flux level
due to any small-scale sky level gradients or very faint diffuse light
features present.  Because the objects we are measuring have such low
surface brightnesses, even small differences in the background flux
levels result in large errors in the colors.  For our example feature
we estimate the uncertainty due to background region selection to be
$\pm\approx0.15$ mag in \BV.  This uncertainty from the selection of
background regions stems primarily from using regions on different
areas of the sky, whereas adjusting the precise borders of a
background region has a comparatively small effect.

The final component of the color uncertainty comes from our treatment
of foreground/background sources which, of course, permeate the image.
Ideally, any method we use to mask these sources should result in a
negligible change in the color since it should affect all regions
equally, and faint sources are expected to be unclustered on these
scales (\eg Brainerd \etal 1995).  We have used two methods to
eliminate the luminosity from bright sources.  In the first, we ran
\emph{objmask} to mask all objects in the image; our final mask
combined all objects detected in either the \V\ or \B\ images.
Alternatively, we used only a $\sigma$-clipping algorithm to remove
high-intensity pixels.  We notice a systematic offset in the colors
using these two masking procedures where the \emph{objmask} technique
produces redder \BV\ colors by $\approx0.1$ mag.  Unfortunately, using
\emph{objmask} to mask sources becomes unfeasible at flux levels much
above 5 ADU ($\approx27.5$ \magsec), as the features we are measuring
are themselves masked.

Each feature's final quoted color range takes into account all of
these effects by making numerous measurements while systematically
varying the input parameters.  While we can estimate the uncertainties
caused by each of these four effects, they are all systematic and
interdependent sources of error, and thus not subject to the usual
methods of error propagation.  We quote the final measured colors as
ranges to emphasize the non-Gaussianity of the errors, especially
given the fact that our results are often not consistent with a
central measured value.

Although this paper is focused almost exclusively on the color of ICL
features, and not their total luminosity or mean surface brightness,
this differential photometry method is also suitable for measuring
those quantities, and mean surface brightness is given for each
feature in the text.  One interesting detail that we find when we
calculate these quantities is that the uncertainty in the color
measurement is smaller than the uncertainty in mean surface brightness
in either band; \ie we can measure the ratio of the fluxes in the two
bands more precisely than we can measure the absolute flux in either
band.  This peculiar effect arises from the fact that the dominant
source of error in this photometric technique is the choice of
background region.  Because the variation between background regions
is caused predominantly by changes in the flux level of astrophysical
backgrounds, such as large-scale ICL components, the flux in these
background regions is correlated in the two bands, and the mean
surface brightness measurements are not independent of one another.
This odd behavior of the photometric uncertainties once again
illustrates that because our measurements are truly limited by
astrophysical backgrounds, the resulting errors are highly systematic
and non-Gaussian.


\begin{thebibliography}{}

\bibitem[Abadi et al.(2006)]{2006MNRAS.365..747A} Abadi, M.~G., Navarro, 
J.~F., \& Steinmetz, M.\ 2006, \mnras, 365, 747 


\bibitem[Adami et 
al.(2005)]{2005A&A...429...39A} Adami, C., et al.\ 2005, \aap, 429, 39 


\bibitem[Aguerri et al.(2005)]{2005AJ....129.2585A} Aguerri, J.~A.~L., 
Gerhard, O.~E., Arnaboldi, M., Napolitano, N.~R., Castro-Rodriguez, N., 
\& Freeman, K.~C.\ 2005, \aj, 129, 2585 


\bibitem[Arnaboldi et al.(2004)]{2004ApJ...614L..33A} Arnaboldi, M., 
Gerhard, O., Aguerri, J.~A.~L., Freeman, K.~C., Napolitano, N.~R., Okamura, 
S., \& Yasuda, N.\ 2004, \apjl, 614, L33 


\bibitem[Baria et al.(2009)]{2009JApA...30....1B} Baria, P., Brito, W., 
\& Martel, H.\ 2009, Journal of Astrophysics and Astronomy, 30, 1 


\bibitem[Bernardi et al.(2003)]{2003AJ....125.1882B} Bernardi, M., et al.\ 
2003, \aj, 125, 1882 


\bibitem[Bernstein et al.(1995)]{1995AJ....110.1507B} Bernstein,
G.~M., Nichol, R.~C., Tyson, J.~A., Ulmer, M.~P., \& Wittman, D.\
1995, \aj, 110, 1507


\bibitem[Brainerd et al.(1995)]{1995MNRAS.275..781B} Brainerd, T.~G.,
Smail, I., \& Mould, J.\ 1995, \mnras, 275, 781


\bibitem[Bruzual \& Charlot(2003)]{2003MNRAS.344.1000B} Bruzual, G.,
\& Charlot, S.\ 2003, \mnras, 344, 1000


\bibitem[Calc{\'a}neo-Rold{\'a}n et al.(2000)]{2000MNRAS.314..324C} 
Calc{\'a}neo-Rold{\'a}n, C., Moore, B., Bland-Hawthorn, J., Malin, D., 
\& Sadler, E.~M.\ 2000, \mnras, 314, 324 


\bibitem[Cantiello et al.(2005)]{2005ApJ...634..239C} Cantiello, M., 
Blakeslee, J.~P., Raimondo, G., Mei, S., Brocato, E., 
\& Capaccioli, M.\ 2005, \apj, 634, 239 


\bibitem[Carollo et al.(1993)]{1993MNRAS.265..553C} Carollo, C.~M., 
Danziger, I.~J., \& Buson, L.\ 1993, \mnras, 265, 553 


\bibitem[Carter \& Dixon(1978)]{1978AJ.....83..574C} Carter, D., \&
Dixon, K.~L.\ 1978, \aj, 83, 574


\bibitem[Castro-Rodrigu{\'e}z et al.(2009)]{2009A&A...507..621C}
Castro-Rodrigu{\'e}z, N., Arnaboldi, M., Aguerri, J.~A.~L., Gerhard,
O., Okamura, S., Yasuda, N., \& Freeman, K.~C.\ 2009, \aap, 507, 621


\bibitem[Coil et al.(2004)]{2004ApJ...617..765C} Coil, A.~L., Newman, 
J.~A., Kaiser, N., Davis, M., Ma, C.-P., Kocevski, D.~D., 
\& Koo, D.~C.\ 2004, \apj, 617, 765 


\bibitem[Connolly et al.(2002)]{2002ApJ...579...42C} Connolly, A.~J., et 
al.\ 2002, \apj, 579, 42 


\bibitem[Conroy et al.(2007)]{2007ApJ...668..826C} Conroy, C., Wechsler, 
R.~H., \& Kravtsov, A.~V.\ 2007, \apj, 668, 826 


\bibitem[Cortese et al.(2010)]{2010MNRAS.403L..26C} Cortese, L., Bendo, 
G.~J., Isaak, K.~G., Davies, J.~I., \& Kent, B.~R.\ 2010, \mnras, 403, L26 


\bibitem[Da Rocha \& Mendes de Oliveira(2005)]{2005MNRAS.364.1069D} Da
Rocha, C., \& Mendes de Oliveira, C.\ 2005, \mnras, 364, 1069


\bibitem[Da Rocha et al.(2008)]{2008MNRAS.388.1433D} Da Rocha, C., Ziegler, 
B.~L., \& Mendes de Oliveira, C.\ 2008, \mnras, 388, 1433 


\bibitem[Davis et al.(1985)]{1985AJ.....90..169D} Davis, L.~E., Cawson, M., 
Davies, R.~L., \& Illingworth, G.\ 1985, \aj, 90, 169 


\bibitem[de Vaucouleurs(1961)]{1961ApJS....5..233D} de Vaucouleurs,
G.\ 1961, \apjs, 5, 233


\bibitem[Dubinski(1998)]{1998ApJ...502..141D} Dubinski, J.\ 1998, \apj, 
502, 141 


\bibitem[Durrell et al.(2002)]{2002ApJ...570..119D} Durrell, P.~R., 
Ciardullo, R., Feldmeier, J.~J., Jacoby, G.~H., 
\& Sigurdsson, S.\ 2002, \apj, 570, 119 


\bibitem[Faure et al.(2007)]{2007A&A...463..833F} Faure, C., Giraud,
E., Melnick, J., Quintana, H., Selman, F., \& Wambsganss, J.\ 2007,
\aap, 463, 833


\bibitem[Feldmeier et al.(1998)]{1998ApJ...503..109F} Feldmeier, J.~J., 
Ciardullo, R., \& Jacoby, G.~H.\ 1998, \apj, 503, 109 


\bibitem[Feldmeier et al.(2004)]{2004ApJ...615..196F} Feldmeier, J.~J., 
Ciardullo, R., Jacoby, G.~H., \& Durrell, P.~R.\ 2004, \apj, 615, 196 


\bibitem[Feldmeier et al.(2004)]{2004ApJ...609..617F} Feldmeier, J.~J., 
Mihos, J.~C., Morrison, H.~L., Harding, P., Kaib, N., 
\& Dubinski, J.\ 2004, \apj, 609, 617 


\bibitem[Feldmeier et al.(2002)]{2002ApJ...575..779F} Feldmeier, J.~J., 
Mihos, J.~C., Morrison, H.~L., Rodney, S.~A., 
\& Harding, P.\ 2002, \apj, 575, 779 


\bibitem[Ferguson et al.(1998)]{1998Natur.391..461F} Ferguson, H.~C., 
Tanvir, N.~R., \& von Hippel, T.\ 1998, \nat, 391, 461 


\bibitem[Gal-Yam et al.(2003)]{2003AJ....125.1087G} Gal-Yam, A., Maoz, D., 
Guhathakurta, P., \& Filippenko, A.~V.\ 2003, \aj, 125, 1087 


\bibitem[Gerhard et al.(2005)]{2005ApJ...621L..93G} Gerhard, O., Arnaboldi, 
M., Freeman, K.~C., Kashikawa, N., Okamura, S., 
\& Yasuda, N.\ 2005, \apjl, 621, L93 


\bibitem[Gnedin(2003)]{2003ApJ...582..141G} Gnedin, O.~Y.\ 2003, \apj, 582, 
141 


\bibitem[Gonzalez et al.(2005)]{2005ApJ...618..195G} Gonzalez, A.~H., 
Zabludoff, A.~I., \& Zaritsky, D.\ 2005, \apj, 618, 195 


\bibitem[Gonzalez et al.(2000)]{2000ApJ...536..561G} Gonzalez, A.~H., 
Zabludoff, A.~I., Zaritsky, D., \& Dalcanton, J.~J.\ 2000, \apj, 536, 561 


\bibitem[Goudfrooij et al.(1994)]{1994A&AS..104..179G} Goudfrooij, P.,
Hansen, L., Jorgensen, H.~E., Norgaard-Nielsen, H.~U., de Jong, T., \&
van den Hoek, L.~B.\ 1994, \aaps, 104, 179


\bibitem[Gregg \& West(1998)]{1998Natur.396..549G} Gregg, M.~D., \&
West, M.~J.\ 1998, \nat, 396, 549


\bibitem[Guhathakurta \& Tyson(1989)]{1989ApJ...346..773G}
Guhathakurta, P., \& Tyson, J.~A.\ 1989, \apj, 346, 773


\bibitem[Krick \& Bernstein(2007)]{2007AJ....134..466K} Krick, J.~E.,
\& Bernstein, R.~A.\ 2007, \aj, 134, 466


\bibitem[Krick et al.(2006)]{2006AJ....131..168K} Krick, J.~E., Bernstein, 
R.~A., \& Pimbblet, K.~A.\ 2006, \aj, 131, 168 


\bibitem[Krick et al.(2010)]{2010AAS...21537804K} Krick, J., et al.\ 2010, 
Bulletin of the American Astronomical Society, 41, 588 


\bibitem[Lang et al.(2010)]{2010AJ....139.1782L} Lang, D., Hogg, D.~W., 
Mierle, K., Blanton, M., \& Roweis, S.\ 2010, \aj, 139, 1782 


\bibitem[Liu et al.(2005)]{2005AJ....129.2628L} Liu, Y., Zhou, X., Ma, J., 
Wu, H., Yang, Y., Li, J., \& Chen, J.\ 2005, \aj, 129, 2628 


\bibitem[Maoz et al.(2005)]{2005ApJ...632..847M} Maoz, D., Waxman, E., 
\& Loeb, A.\ 2005, \apj, 632, 847 


\bibitem[McGee \& Balogh(2010)]{2010MNRAS.403L..79M} McGee, S.~L., \&
Balogh, M.~L.\ 2010, \mnras, 403, L79


\bibitem[Mihos et al.(2005)]{2005ApJ...631L..41M} Mihos, J.~C., Harding, 
P., Feldmeier, J., \& Morrison, H.\ 2005, \apjl, 631, L41 


\bibitem[Monaco et al.(2006)]{2006ApJ...652L..89M} Monaco, P., Murante, G., 
Borgani, S., \& Fontanot, F.\ 2006, \apjl, 652, L89 


\bibitem[Moore et al.(1996)]{1996Natur.379..613M} Moore, B., Katz, N., 
Lake, G., Dressler, A., \& Oemler, A.\ 1996, \nat, 379, 613 


\bibitem[Morganson 
\& Blandford(2009)]{2009MNRAS.398..769M} Morganson, E., \& Blandford, R.\ 2009, \mnras, 398, 769 


\bibitem[Murante et al.(2004)]{2004ApJ...607L..83M} Murante, G., et al.\ 
2004, \apjl, 607, L83 


\bibitem[Murante et al.(2007)]{2007MNRAS.377....2M} Murante, G., Giovalli, 
M., Gerhard, O., Arnaboldi, M., Borgani, S., 
\& Dolag, K.\ 2007, \mnras, 377, 2 


\bibitem[Napolitano et al.(2003)]{2003ApJ...594..172N} Napolitano, N.~R., 
et al.\ 2003, \apj, 594, 172 


\bibitem[Neill et al.(2005)]{2005ApJ...618..692N} Neill, J.~D., Shara, 
M.~M., \& Oegerle, W.~R.\ 2005, \apj, 618, 692 


\bibitem[Pettit(1954)]{1954ApJ...120..413P} Pettit, E.\ 1954, \apj,
120, 413


\bibitem[Pierini et 
al.(2008)]{2008A&A...483..727P} Pierini, D., Zibetti, S., Braglia, F., B{\"o}hringer, H., Finoguenov, A., Lynam, P.~D., \& Zhang, Y.-Y.\ 2008, \aap, 483, 727 


\bibitem[Purcell et al.(2008)]{2008MNRAS.391..550P} Purcell, C.~W., 
Bullock, J.~S., \& Zentner, A.~R.\ 2008, \mnras, 391, 550 


\bibitem[Purcell et al.(2007)]{2007ApJ...666...20P} Purcell, C.~W., 
Bullock, J.~S., \& Zentner, A.~R.\ 2007, \apj, 666, 20 


\bibitem[Rawle et al.(2010)]{2010MNRAS.401..852R} Rawle, T.~D., Smith, 
R.~J., \& Lucey, J.~R.\ 2010, \mnras, 401, 852 


\bibitem[Rudick et al.(2006)]{2006ApJ...648..936R} Rudick, C.~S., Mihos, 
J.~C., \& McBride, C.\ 2006, \apj, 648, 936 


\bibitem[Rudick et al.(2009)]{2009ApJ...699.1518R} Rudick, C.~S., Mihos, 
J.~C., Frey, L.~H., \& McBride, C.~K.\ 2009, \apj, 699, 1518 

\bibitem[Ruszkowski 
\& Springel(2009)]{2009ApJ...696.1094R} Ruszkowski, M., \& Springel, V.\ 2009, \apj, 696, 1094 


\bibitem[S{\'a}nchez-Bl{\'a}zquez et al.(2007)]{2007MNRAS.377..759S}
S{\'a}nchez-Bl{\'a}zquez, P., Forbes, D.~A., Strader, J., Brodie, J.,
\& Proctor, R.\ 2007, \mnras, 377, 759


\bibitem[Sandage(1976)]{1976AJ.....81..954S} Sandage, A.\ 1976, \aj, 81, 
954 


\bibitem[Schlegel et al.(1998)]{1998ApJ...500..525S} Schlegel, D.~J., 
Finkbeiner, D.~P., \& Davis, M.\ 1998, \apj, 500, 525 


\bibitem[Slater et al.(2009)]{2009PASP..121.1267S} Slater, C.~T., Harding, 
P., \& Mihos, J.~C.\ 2009, \pasp, 121, 1267 


\bibitem[Sommer-Larsen et al.(2005)]{2005MNRAS.357..478S} Sommer-Larsen, 
J., Romeo, A.~D., \& Portinari, L.\ 2005, \mnras, 357, 478 


\bibitem[Spinrad et al.(1972)]{1972ApJ...175..649S} Spinrad, H.,
Smith, H.~E., \& Taylor, D.~J.\ 1972, \apj, 175, 649


\bibitem[Strom et al.(1976)]{1976ApJ...204..684S} Strom, S.~E., Strom,
K.~M., Goad, J.~W., Vrba, F.~J., \& Rice, W.\ 1976, \apj, 204, 684


\bibitem[Strom \& Strom(1978)]{1978AJ.....83...73S} Strom, K.~M., \&
Strom, S.~E.\ 1978, \aj, 83, 73


\bibitem[Tamura et al.(2000)]{2000AJ....119.2134T} Tamura, N., Kobayashi, 
C., Arimoto, N., Kodama, T., \& Ohta, K.\ 2000, \aj, 119, 2134 


\bibitem[Trentham \& Mobasher(1998)]{1998MNRAS.293...53T} Trentham,
N., \& Mobasher, B.\ 1998, \mnras, 293, 53


\bibitem[Uson et al.(1991)]{1991ApJ...369...46U} Uson, J.~M., Boughn,
S.~P., \& Kuhn, J.~R.\ 1991, \apj, 369, 46


\bibitem[Vader et al.(1988)]{1988A&A...203..217V} Vader, J.~P.,
Vigroux, L., Lachieze-Rey, M., \& Souviron, J.\ 1988, \aap, 203, 217


\bibitem[van Zee et al.(2004)]{2004AJ....128.2797V} van Zee, L., Barton, 
E.~J., \& Skillman, E.~D.\ 2004, \aj, 128, 2797 


\bibitem[Vilchez-Gomez et al.(1994)]{1994A&A...283...37V}
Vilchez-Gomez, R., Pello, R., \& Sanahuja, B.\ 1994, \aap, 283, 37


\bibitem[Villumsen et al.(1997)]{1997ApJ...481..578V} Villumsen, J.~V., 
Freudling, W., \& da Costa, L.~N.\ 1997, \apj, 481, 578 


\bibitem[White et al.(2003)]{2003ApJ...585..739W} White, P.~M., Bothun, G., 
Guerrero, M.~A., West, M.~J., \& Barkhouse, W.~A.\ 2003, \apj, 585, 739 


\bibitem[Williams et al.(2007)]{2007ApJ...656..756W} Williams, B.~F., et 
al.\ 2007, \apj, 656, 756 


\bibitem[Willman et al.(2004)]{2004MNRAS.355..159W} Willman, B., Governato, 
F., Wadsley, J., \& Quinn, T.\ 2004, \mnras, 355, 159 


\bibitem[Witt et al.(2008)]{2008ApJ...679..497W} Witt, A.~N., Mandel,
S., Sell, P.~H., Dixon, T., \& Vijh, U.~P.\ 2008, \apj, 679, 497


\bibitem[Yagi et al.(2007)]{2007ApJ...660.1209Y} Yagi, M., Komiyama,
Y., Yoshida, M., Furusawa, H., Kashikawa, N., Koyama, Y., \& Okamura,
S.\ 2007, \apj, 660, 1209


\bibitem[Zeilinger et al.(1993)]{1993MNRAS.261..175Z} Zeilinger,
W.~W., Moller, P., \& Stiavelli, M.\ 1993, \mnras, 261, 175


\bibitem[Zibetti et al.(2005)]{2005MNRAS.358..949Z} Zibetti, S., White, 
S.~D.~M., Schneider, D.~P., \& Brinkmann, J.\ 2005, \mnras, 358, 949 


\bibitem[Zwicky(1951)]{1951PASP...63...61Z} Zwicky, F.\ 1951, \pasp, 63, 61 


\end{thebibliography}
\end{document}